\documentclass[preprint,showkeys,showpacs]{revtex4-1}
\usepackage{bm}
\usepackage{multirow}
\usepackage[dvips]{graphicx}
\usepackage{amsmath}
\usepackage[rflt]{floatflt}
\usepackage{booktabs}
\usepackage{array}

\newcommand{\mieq}[1]{eq. \eqref{#1}}
\newcommand{\mitable}[1]{Table \ref{#1}}
\begin{document}

\title{The modified Becke-Johnson potential analyzed.}
\author{J.A. Camargo-Mart\'inez}
\affiliation{Departamento de F\'isica, CINVESTAV-IPN, Av. IPN 2508, 07360 M\'exico}
\author{R. Baquero}
\affiliation{Departamento de F\'isica, CINVESTAV-IPN, Av. IPN 2508, 07360 M\'exico}

\begin{abstract}
Recently in the Wien2k code, the modified Becke-Johnson potential (mBJLDA) was implemented. As the authors [{\em Phys.Rev.Lett.} 102, 226401 (2009)] point,
this potential reproduces the band gap of semiconductors with improved accuracy. In this paper we present our analysis of this potential in two directions.
First, we checked whether this potential reproduces the band structure for metals, an analysis that lacked in the literature. We calculated the band gap of a group 
of semiconductors. We observed that the Linear Density Approximation (LDA) give rise to a shorter lattice constant as compared to experiment. The Generalized
Gradient Approximation behaves oppositely. Using the average, $a_{Avg}$, in the mBJLDA potential, we obtained a closer to experiment value for the gap. 
We conclude that the new mBJLDA potential represent an important improvement as compared to the results from the previous version of the Wien2k code. Also the
mBJLDA potential can be a very useful tool for the theoretical study of complex systems containing semiconductor compounds such as surfaces, superlattices and interfaces. 
\end{abstract}



\maketitle

\section{Introduction \label{introduction}}
Density Functional Theory (\emph{DFT}) is nowadays the most used method to calculate band structures. It is implemented in several codes.
One well known is the Wien2k code. It has evolved in several versions. A friendly interaction was produced already in the 
$2000$ version. Nevertheless, a long standing problem of all the codes based on \emph{DFT} was that the band structure of semiconductors, in spite of giving a 
reasonable account of the dispersion of the bands, was systematically unable to reproduce the experimental values of the gap, This problem 
could be solved by hand using a “trick” but this kind of solution is not what we expect from an “ab initio” calculation. 
The removal of this problem in the new version (Wien2k 2011) is the subject of the analysis that we present in this work. 
The new ability to reproduce the gap value of semiconducting materials, allows confirming results for semiconductor/metal 
and semiconductor/semiconductor interfaces and to calculate new ones. These results are of technological interest and per se. 
For example, the \emph{YBCO7/GaAs}($001$) interface was calculated \cite{rafael} using the previous version of the Wien2k code ($2008$ version) and 
two atomic planes in the GaAs side of the interface were found to be metallic.  This result could, nevertheless, be influenced by 
the inability of the previous version of the code to properly account for the gap of the semiconductor. Since the “trick” mentioned 
above could not be used in an interface calculation, the result remains questionable because of the uncertainties in the \emph{GaAs} side
results around the very important gap region. The rest of the paper is organized as follows. In the next section II, we deal very 
briefly with Density Functional theory (\emph{DFT}) to point to a detail important to this work. In section III, we briefly present a few 
of our calculated results for the band structure of metals to check that the accuracy of the new version remains the same also in 
this case.  Section IV is devoted to semiconductors. We recalculate the results from references \cite{blaha} and \cite{hse} and present new results 
for some other semiconductors which we compare to experiment. In a final section V, we analyze the results with the new code and 
present our conclusions.

\section{Density Functional Theory }
In solids, ions and electrons constitute a many body interacting system described by a Schrödinger equation with too many particle
coordinates to be numerically treatable nowadays, as it is very well known. During the last decade, several codes were developed 
based on \emph{DFT} and this method became the most used, precise and practical way to calculate the band structure 
of solids. The development of practical approximations to the correlation and interchange potential lead to a remarkable degree of 
accuracy to describe even complicated metallic systems. At the basis of \emph{DFT} is the celebrated Hohenberg-Khon theorem which shows that 
the density of the ground state contains all the possible information on a system and its knowledge is equivalent to the wave function
itself. So, the expectation value~\cite{4} of any observable can be calculated from a unique functional of the ground state density, $\rho(r)$, 
which minimizes the energy functional, E[$\rho$]. Further, Khon and Sham~\cite{5} transformed the many-body problem into a one-body problem and 
showed that the density of states calculated from the solution of the so-called Khon-Sham equations~\eqref{eq1} is equal to the one of the 
real ground state density of the many-body system,
\begin{equation}
[T+V_{H}+V_{ext}+V_{xc}]\varphi_{i} (r)= \varepsilon_{i}\varphi_{i}
\label{eq1}
\end{equation}
Where the density is calculated taking into account the occupied states only. In \mieq{eq1}, T is the kinetic energy operator, $V_{H}$ is 
the Hartree potential and $V_{xc}$ is the exchange and correlation potential which is calculated from the exchange and correlation 
energy functional,$V_{xc}(r)= \frac{\delta E_{xc}[\rho]}{\delta\rho}.$ To solve the Khon-Sham equations~\ref{eq1}, an explicit expression for 
$E_{xc}[\rho]$ is needed. The exact expression is unknown since it includes all kind of correlations between all the particles in the system. 
So an approximation is needed. The first and best known approximation is the Local Density Approximation, \emph{LDA}~\cite{6}, which was followed by the 
Generalized Gradient Approximation (\emph{GGA})~\cite{6} and the \emph{meta-GGA}~\cite{7} among other approximations. These potentials reproduce rather well the 
band structure of even complicated metallic systems but fail in reproducing the gap in semiconductors. A recent progress has been made. 
Blaha et al.~\cite{blaha} have reported the so-called mBJLDA potential which is a modification of the exchange and correlation potential of
Becker and Johnson (\emph{BJ})~\cite{8}. The new potential reproduces the experimental gaps of semiconductors with an accuracy several orders of magnitude better 
than the former existing potentials. The modified mBJLDA potential is 
\begin{equation}
V_{x,\sigma}^{mBJ}(r)=cV_{x,\sigma}^{BR}(r)+(3c-2)\frac{1}{\pi}\frac{\sqrt{5}}{12}\frac{\sqrt{2t_{\sigma}(r)}}{\rho_\sigma(r)}
\label{eq2}
\end{equation}
Where $\rho_\sigma(r)$ is the density of states, $t_\sigma(r)$ is the kinetic energy density and $V_{x,\sigma}^{BR}(r)$ is the Becke-Roussel
potential (\emph{BR})~\cite{9}. The c stands for,
\begin{equation}
c=\alpha + \left(\beta\frac{1}{V_{cell}}\int{d^3 r\frac{\mid \nabla \rho(r)\mid}{\rho(r)}}\right)^{1/2}
\label{eq3}
\end{equation}
$\alpha$ and $\beta$ are free parameters. Within the Wien2k code~\cite{10} $\alpha=-0.012$ and $\beta=1.023$ $Bohr^{1/2}$. 

\section{Metal calculations}
We present here our result for Nb, V and Ta, to check whether differences arise between the LDA and the new mBJLDA potential
for the case of metal. This analysis does not appear reported in the literature, and are important to the calculation of metal/semiconductor interfaces.
We have calculated  these band structures, first, using the LDA approximation with the old Wien2k code and then we redid the same calculation using the new mBJLDA potential.
To optimize the lattice parameter in a consistent way it is recommended \cite{blaha} to use \emph{LDA} (\emph{GGA}) first and to use further the optimized 
lattice parameter obtained in this way to compute the band gap structure with the mBJLDA potential. If we follow this method, we find a good agreement between 
\emph{LDA} and mBJLDA for all the three metals calculated as it can be checked from \mitable{T1}. 
 
\begin{table}[h]
\caption{The lattice parameter, a, in Angstrom; the Fermi energy, $E_F$, in Rydbergs; the density of states at $E_F$, 
$N(E_F)$ in states per Rydberg for Nb, V and Ta calculated with LDA and with mBJLDA potential. The experimental values were taken from reference \cite{11}.}
\begin{tabular*}{0.65\textwidth}{ccccccc}\hline\hline
        & Experiment & \multicolumn{3}{c}{LDA} & \multicolumn{2}{c}{mBJLDA}\\\hline
Element &      a     &     a    &   $E_F$   & $N(E_F)$ &   $E_F$  & $N(E_F)$  \\\hline
Nb      &   3.30059  &  3.2487  &  0.78890  &    24    &  0.7785  &   22.15  \\
V       &   3.02487  &  2.9273  &  0.67200  &  28.28   &  0.6734  &   29.24   \\
Ta      &   3.30280  &  3.2500  &  0.84131  &  21.20   &  0.8572  &   20.78   \\
\hline\hline
\end{tabular*}
\label{T1}
\end{table}

The Fermi energy values, $E_F$, calculated with the mBJLDA potential, for \emph{Nb}, differs in 0.01 Ry with respect to the \emph{LDA} value, which 
represent a difference of 1.3\%. For \emph{V} and \emph{Ta} these values are 0.2\% and 1.8\% respectively. The density of states at $E_F$, 
$N(E_F)$, presents difference of 7.7\%, 3.4\% and 2.0\% for \emph{Nb}, \emph{V} and \emph{Ta} respectively. 
We omit the plot of the band structure and the density of states obtained in the different ways mentioned here since the overall agreement 
is such that the details just discussed do not show explicitly enough and these band structures are very well known. For \emph{Nb} we have compared 
our results with references \cite{12,13,14}, \emph{V} with references \cite{14,15} and for \emph{Ta} with reference \cite{16}.
These results show that the mBJLDA potential reproduces well the band structure of metals.

\section{Semiconductor calculations}

A particular feature of mBJLDA potential is that a corresponding exchange and correlation energy term , $E_{xc}[\rho]$, such that the mBJLDA potential is obtained 
in the usual way, namely, $V_{xc}=\delta E_{xc}[\rho]/\delta \rho$, is not possible. As a consequence, a consistent optimization procedure to obtain the lattice 
parameter, the Bulk modulus and its derivative with respect to pressure are not actually possible. This is a consequence of the empirical character of this 
potential. For that reason, Tran and Blaha have proposed the empirical alternative that prior to a band structure calculation with the mBJLDA potential, the 
lattice parameter is found from either a LDA or a GGA optimization procedure and the result introduced into the code to perform the band structure calculation of 
the semiconductor system. Such a procedure gives rise to quite improved results as compared to the previous version of the Wien2k code, as we stated before. It is known that the LDA 
underestimates as a rule, the lattice parameters and, on the contrary, GGA overestimates them. We have explored the possibility of using the averaged value as the 
lattice parameter, $a_{Avg}$, where $a_{Avg}=(a_{LDA}+a_{GGA})/2$. Here  $a_{LDA}(a_{GGA}$) is the lattice parameter obtained from an LDA (GGA) optimization 
procedure. When $a_{Avg}$ is used as input into the Wien2k code implemented with the mBJLDA potential, a better agreement of the band gap value with experiment is 
obtained as compared to the results with $a_{LDA}$. So this procedure turns out to give better results than the one recommended by Tran and Blaha and its extra 
computational cost is relatively low. 

In \mitable{T3}, we present the gap value obtained from our band structure calculations for several semiconductors using \emph{LDA} and the new mBJLDA potential,
using as lattice parameters $a_{LDA}$ and $a_{Avg}$, called mBJLDA($a_{LDA}$) and mBJLDA($a_{Avg}$) respectively, and compare our results with the ones reported 
by Blaha et al. \cite{blaha}, with the ones obtained using  

\begin{table}[h]
\caption{\label{T3}The gap is in eV, the crystal structure  is indicated in the second column, the data are from Blaha et al. \cite{blaha}, 
from HSE \cite{hse} and the experimental ones from references \cite{blaha,hse,17,18}. The absolute percentage error with respect to experiment 
is shown in parentheses.}
\begin{tabular*}{1\textwidth}{@{\extracolsep{\fill}}cc|ccc|c|c|c}\hline\hline
        &           &                &       This work             &                   &            Blaha et al.          &         &        \\\hline
Element & Structure &   LDA          &   mBJLDA($a_{LDA}$)         & mBJLDA($a_{Avg}$) &            mBJLDA                &   HSE   &  Expt. \\\hline
Si      &     A1    &   0.48 (59\%)  &        1.13 (3.4\%)         & 1.17 (0.0\%)      &             1.17 (0.0\%)         &   1.28  &  1.17  \\
Ge      &     A1    &   0.00 (100\%) &        0.91 (23.0\%)        & 0.80 (8.1\%)      &             0.85 (14.9\%)        &   0.56  &  0.74  \\
MgO     &     B1    &   4.72 (38\%)  &        7.57 (0.3\%)         & 7.22 (4.4\%)      &             7.17 (5.0\%)         &   6.50  &  7.55  \\
LiF     &     B1    &   8.78 (38\%)  &        13.8 (2.8\%)         & 13.4 (5.6\%)      &             12.9 (8.9\%)         &         &  14.20 \\
AlAs    &     B3    &   1.35 (39\%)  &        2.13 (4.5\%)         & 2.17 (2.7\%)      &                                  &   2.24  &  2.23  \\
SiC     &     B3    &   1.31 (45\%)  &        2.21 (7.9\%)         & 2.26 (5.8\%)      &             2.28 (5.0\%)         &   2.39  &  2.40  \\
BP      &     B3    &   1.19 (40\%)  &        1.80 (10.0\%)        & 1.83 (8.5\%)      &                                  &   2.16  &  2.00  \\
BAs     &     B3    &   1.23 (16\%)  &        1.69 (15.8\%)        & 1.72 (17.8\%)     &                                  &   1.92  &  1.46  \\
InP     &     B3    &   0.45 (69\%)  &        1.70 (18.9\%)        & 1.52 (6.3\%)      &       1.40$^\text{a}$ (14.7\%)   &   1.64  &  1.43  \\
AlP     &     B3    &   1.45 (41\%)  &        2.28 (6.9\%)         & 2.33 (4.9\%)      &             2.32 (5.3\%)         &   2.52  &  2.45  \\
BN      &     B3    &   4.78 (23\%)  &        5.86 (5.8\%)         & 5.85 (5.9\%)      &             5.85 (5.9\%)         &   5.98  &  6.22  \\
GaN     &     B3    &   1.66 (48\%)  &        3.13 (2.2\%)         & 2.94 (8.1\%)      &             2.81 (12.2\%)        &   3.03  &  3.20  \\
CdTe    &     B3    &   0.49 (67\%)  &        1.80 (20.8\%)        & 1.67 (12.1\%)     &                                  &   1.52  &  1.49  \\
GaAS    &     B3    &   0.30 (80\%)  &        1.84 (21.1\%)        & 1.56 (2.6\%)      &             1.64 (7.9\%)         &   1.21  &  1.52  \\
ZnS     &     B3    &   1.85 (53\%)  &        3.63 (7.2\%)         & 3.70 (5.4\%)      &             3.66 (6.4\%)         &   3.42  &  3.91  \\
CdS     &     B3    &   0.87 (64\%)  &        2.68 (10.7\%)        & 2.61 (7.9\%)      &             2.66 (9.9\%)         &   2.14  &  2.42  \\
AlSb    &     B3    &   1.14 (32\%)  &        1.76 (4.8\%)         & 1.80 (7.1\%)      &                                  &   1.99  &  1.68  \\
InN     &     B4    &   0.02 (97\%)  &        0.82 (18.8\%)        & 0.82 (18.8\%)     &                                  &   0.71  &  0.69  \\
AlN     &     B4    &   4.14 (34\%)  &        5.52 (12.1\%)        & 5.53 (11.9\%)     &             5.55 (11.6\%)        &         &  6.28  \\
ZnO     &     B4    &   0.75 (78\%)  &        2.77 (18.5\%)        & 2.76 (19.8\%)     &             2.68 (22.1\%)        &         &  3.44  \\\hline
\multicolumn{2}{l}{Average error}        &     53\%       &           10.8\%            &   8.2\%           &                  9.3\%           &         &        \\
\hline\hline
\end{tabular*}
\footnotetext{Reference \cite{18}.}
\end{table}
\newpage
\clearpage

the hybridized exchange potential of \emph{Heyd-Scuseria-Ernzerhof} (\emph{HSE}) reported in reference \cite{hse} and to the experimental values reported in 
references \cite{blaha,hse,17,18}.

It is evident from \mitable{T3} that the gap values for semiconducting systems calculated with \emph{LDA} turn out to be wrong as it is very well known. We can 
see clearly that the results obtained using the mBJLDA potential shows a significant improvement in the calculation of the gap with respect to experiment.
Now,the values calculated with mBJLDA($a_{Avg}$) have a significantly better agreement with experiment of the gap as compared to the one obtained with mBJLDA($a_{LDA}$) and 
with the values reported by Blaha. The average absolute error values for each procedure were 8.2\% for mBJLDA($a_{Avg}$), 9.3\% for values reported by Blaha and 10.8\% for 
mBJLDA($a_{LDA}$). The results calculated with the \emph{HSE} (See \mitable{T3}) present a good agreement with experiment too. We conclude that our proposal to use the average 
value, $a_{Avg}$, for calculating the band gap using the mBJLDA potential in the Wien2k code results in better agreement with the experimental values. The new mBJLDA potential opens 
the possibility to carry out theoretical studies of complex systems containing semiconductor compounds as surfaces, superlattices and interfaces. 

\section{Conclusions}
In this work we performed an analysis of the new progress done in the implementation of \emph{DFT} whose most famous shortcome was its impossibility to account 
for the experimental value of the band gap of semiconducting systems. We have calculated using the new mBJLDA potential~\cite{blaha}, the band structure of some
semiconductors and got their band gap value which we compared to experiment.
In this work, we found two important facts. First, the mBJLDA potential reproduces correctly the band structure of metals, which is an important new observation.
This result is not reported in the literature. Second, that the best result for the band gap value is obtained, in general, if the average 
lattice parameter, $a_{Avg}$, is used ($a_{Avg}=(a_{LDA}+a_{GGA})/2$) where $a_{LDA}$($a_{GGA}$) is the lattice parameter that results from a LDA(GGA) optimization, 
We have calculated the band gap for all semiconductors reported in ref.~\cite{blaha} and ref.~\cite{hse} and some other and compare the results among themselves 
and with experiment. We found that the new mBJLDA potential gives rise to gap values that represent an important progress as compare to the 
old LDA potential. This new mBJLDA potential, allows the calculation of interfaces and superlattices with a semiconducting component 
with a high degree of accuracy which were difficult with the old code. For example, in the ref.~\cite{rafael} the electronic band structure of the interface \emph{YBCO7/GaAs}
was calculated using Wien2k 2008 code and obtained that two atomic planes in the GaAs side become metallic. 
Nevertheless since the code does not allow the correct calculation of the gap of the semiconductor, this interesting result remained 
uncertain~\cite{raul}. In that sense, the new potential opens a new field which is of interest in several disciplines as spintronics, semiconductor devices, 
superconductivity or two-dimensional electron gas properties, among others.

\section*{Acknowledgments}

The authors acknowledge to the GENERAL COORDINATION OF INFORMATION AND COMMUNICATIONS TECHNOLOGIES (CGSTIC) at CINVESTAV for providing HPC resources on the Hybrid Cluster Supercomputer "Xiuhcoatl", that have
contributed to the research results reported within this paper.

\end{document}